\documentclass[twocolumn,amsmath,amssymb,floatfix,prl,showpacs,footinbib]{revtex4}
\usepackage{amsmath,amssymb,natbib,bm,graphicx,url,psfrag}

\usepackage{color}

\newcommand{\bra}[1]{\left\langle \, #1 \,\right|}
\newcommand{\ket}[1]{\left|\, #1 \, \right\rangle}

\newcommand{\be}{\begin{equation}}
\newcommand{\ee}{\end{equation}}

\begin{document}

\title{Discontinuous current-phase relations in small 1D Josephson junction arrays}
\author{Jens Koch}
\affiliation{Departments of Physics and Applied Physics, Yale University, PO Box 208120, New Haven, CT 06520, USA}
\author{Karyn Le Hur}
\affiliation{Departments of Physics and Applied Physics, Yale University, PO Box 208120, New Haven, CT 06520, USA}

\begin{abstract}
We study the Josephson effect in small one-dimensional (1D) Josephson junction arrays. For weak Josephson tunneling, topologically different regions in the charge-stability diagram generate distinct current-phase relationships (I$\Phi$s). We present results for a three-junction system in the vicinity of charge degeneracy lines and triple points. We explain the generalization to larger arrays, show that discontinuities of the I$\Phi$ at phase $\pi$ persist and that, at maximum degeneracy, the problem can be mapped to a tight-binding model providing analytical results for arbitrary system size.
\end{abstract}

\date{February 16, 2008} 
\maketitle

\begin{figure}
 \centering
 \includegraphics[width=1.0\columnwidth]{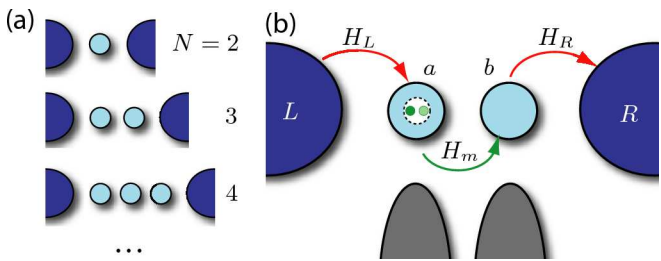}
 \caption{(Color online) (a) Progression of small 1D Josephson junction arrays with increasing number of junctions $N$. (b) Detailed schematic of the three-junction JJA. Electronic transport across the junctions is due to inter-dot ($H_m$) and lead-dot ($H_{L,R}$) tunneling.
 The electrochemical potentials of the dots can be adjusted  by separate gate electrodes.
 \label{fig:fig1}}
 \end{figure}

\emph{Introduction}.---The Josephson current-phase relation (I$\Phi$) contains information about the microscopic nature of the Cooper pair transfer between superconductors. Specifically, systems more complicated than a single tunnel junction generally exhibit non-sinusoidal I$\Phi$s \cite{golubov}. A pertinent example consists of the double junction formed by one superconducting grain coupled to two leads \cite{averin1991,1993PhRvL..70.2940M}. In this Letter, we go beyond the single-grain case and investigate the Josephson effect in small 1D Josephson junction arrays (JJAs), cf.\ Fig.\ \ref{fig:fig1}(a).

As demonstrated for the two-junction case, charging effects can play an important role in the transport of Cooper pairs due to the competition between Coulomb blockade and tunneling of charge. This can significantly modify the I$\Phi$ and the critical current \cite{averin1991,1993PhRvL..70.2940M}.  To be precise, the I$\Phi$ of the two-junction system differs maximally from a sinusoid at the charge degeneracy point. Given that both junctions are equally strong, the I$\Phi$ is in fact discontinuous at $\phi=\pi$, where $\phi$ denotes the gauge-invariant phase difference across the entire array.

These results are obtained in the particularly interesting regime where superconductivity is fully developed ($T\ll E_C \ll \Delta$, the respective quantities denoting the temperature, relevant charging energy, and superconducting gap), but charging effects are strong compared to Josephson tunneling, $E_C\gg E_J$. In this case, the charge-stability diagram \cite{wiel} is an appropriate starting point for identifying the relevant states, and for classifying the different types of I$\Phi$s. For instance, the charge-stability diagram of the two-junction system consists of isolated points along the gate-voltage axis, specifiying the potentials where two charge states become energetically degenerate. In the vicinity of these charge-degeneracy points, two charge states participate in the Cooper pair transfer and the I$\Phi$ is strongly non-sinusoidal. Away from the degeneracy points, the system remains in a single charge state and exhibits a sinusoidal I$\Phi$.

This reasoning can be extended to 1D JJAs with larger numbers of grains, making the connection between the limiting cases of a single grain and the limit of an infinite array. It is important to note, however, that the regime considered here is different from the self-charging and nearest-neighbor models for infinite 1D JJAs \cite{doniach,choi93,matveev2002}. In our case, the total number of junctions in the array determines the maximum number of charge states which can become degenerate. The maximal degeneracy is generally represented by a point in the charge-stability diagram. Lower-order degeneracies are associated with hyperplanes of different dimensionalities (lines and planes being simple examples), generating distinct I$\Phi$s. The accessibility of such degeneracies up to fourth order has recently been demonstrated in normal-state triple dots \cite{schroer}. We show that discontinuities in I$\Phi$ persist beyond the single-grain case. We  present a detailed analysis of the three-junction system shown in Fig.\ \ref{fig:fig1}(b), and then describe the generalization to larger systems; at maximum degeneracy we obtain  a useful tight-binding approach.

\emph{Model}.---For three junctions, the system consists of two superconducting quantum dots $a$ and $b$ in series with two macroscopic superconducting leads $L$ and $R$, closely resembling devices previously studied in various contexts in experiments \cite{geerligs,pashkin2003,bibow}.  The dots and leads are coupled through Josephson junctions with Josephson energies $E_{J\alpha}$ and junction capacitances $C_\alpha$, where $\alpha=L,R,m$ signifies the left, right, and middle junction. Additionally, the two dots are capacitively coupled to separate gate electrodes, enabling the individual tuning of their electrochemical potentials. 

\begin{figure}
\includegraphics[width=0.95\columnwidth]{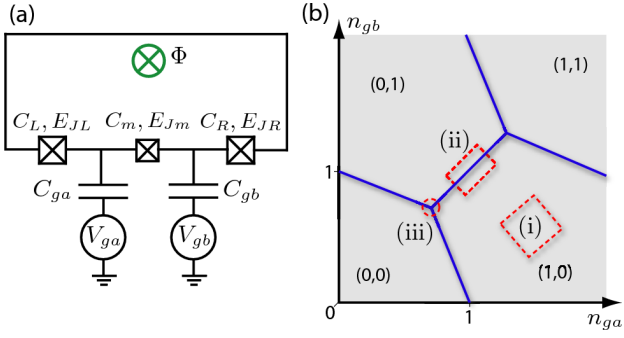} 
 \caption{(color online) 
(a) Lumped element circuit for the superconducting double quantum dot. The supercurrent in the loop is induced by applying an external magnetic flux $\Phi$ and can be measured by an inductively coupled tank circuit (not shown).
(b) Charge-stability diagram for the superconducting double-dot system. The dashed regions mark the regimes (i)--(iii) away from all degeneracies, close to a charge-degeneracy line, and close to a triple point of the honeycomb pattern.
}
\label{fig:fig2}
 \end{figure}

We focus on the scenario where the superconductivity is fully developed, captured by BCS theory, and not impeded by charging effects ($T\ll E_C\ll\Delta$). In this case, breaking of Cooper pairs due to thermal fluctuations or charging effects is negligible, and the system Hamiltonian can be derived in a convenient way via circuit quantization \cite{devoret2}, cf.\ Fig.\ \ref{fig:fig2}(a).
The Hamiltonian consists of charging and Josephson tunneling contributions, $\hat{H}=\hat{H}_\text{ch}+\sum_\alpha\hat{H}_\alpha$. The first term describes the charging energy of the double-dot system \cite{pashkin2003,li:176802},
\begin{align}
\hat{H}_\text{ch}=&\sum_{j=a,b}E_{Cj}(\hat{n}_j-n_{gj})^2+ E_m\prod_{j=a,b}(\hat{n}_j-n_{gj}),\label{hch}
\end{align}
where $\hat{n}_a$, $\hat{n}_b$ count the number of extra Cooper pairs on dots $a$ and $b$, respectively.
For convenience, we have introduced the abbreviations
\be
E_{Ca,b} = \frac{(2e)^2 C_{\Sigma b,a}}{2(C_{\Sigma a}C_{\Sigma b}-C_m^2)},\,\,
E_m = \frac{(2e)^2 C_m}{C_{\Sigma a}C_{\Sigma b}-C_m^2}
\ee
for the partial charging energies. The tunable offset charges are denoted by  $n_{g\alpha}=C_{g\alpha}V_{g\alpha}/2e$, and $C_{\Sigma a,b}=C_m+C_{L,R}+C_{ga,b}$. In the following, we will restrict to the case where all junction capacitances are finite. The corresponding charge-stability diagram  is shown in Fig.\ \ref{fig:fig2}(b). For finite $C_m$, it forms a honeycomb pattern typical of double dots \cite{wiel}, and features three distinct regimes: (i) regions away from any charge degeneracies, (ii) charge degeneracy lines, and (iii) triple points.

The tunneling of Cooper pairs is captured by the Josephson terms $\hat{H}_\alpha$, where
\begin{align}\nonumber
\hat{H}_L &= -E_{JL}\cos (\varphi_L-\hat{\varphi}_a),\quad
\hat{H}_R = -E_{JR}\cos (\varphi_R-\hat{\varphi}_b),\\
\hat{H}_m &= -E_{Jm}\cos (\hat{\varphi}_a-\hat{\varphi}_b).
\end{align}
Here, the quantities $\varphi_j$ with $j=a,b,L,R$ denote the phases used to build the gauge-invariant phase differences across the junctions. Note that the leads $L$ and $R$ are assumed to be macroscopic superconductors with negligible charging energy. As a result, as opposed to the phases on the grains their phases may be treated  ``classically".

\emph{Current-phase relations}.---%
The I$\Phi$ may be probed in a loop configuration as shown in Fig.\ \ref{fig:fig2}(a), employing the ``rf technique" discussed in Ref.\ \cite{golubov}.  Flux quantization relates the external flux $\Phi$ to the phase difference $\phi$ via $\phi=\varphi_L-\varphi_R=2\pi\Phi/\Phi_0$, where $\Phi_0=h/2e$ denotes the flux quantum.
Based on the above model, we evaluate the I$\Phi$s in the regimes (i)--(iii) determined by the charge-stability diagram. For (i), i.e.\ away from charge degeneracies, it is straightforward to see that each Cooper pair must coherently tunnel from one lead to the other, occupying the grains only in virtual intermediate states. The result is a sinusoidal I$\Phi$, analogous to the single-grain case away from charge degeneracy. Here, however, the critical current is suppressed by a factor $(E_{J\alpha}/E_C)^2$.

Close to charge-degeneracy lines [regime (ii)], two charge states participate in the Josephson tunneling. Without loss of generality, we may focus on the gate-voltage region marked in Fig.\ \ref{fig:fig2}(b), where the charge states $(1,0)$ and $(0,1)$ with an additional Cooper pair on either dot $a$ or $b$ are close to degeneracy. Our assumption of  weak Josephson tunneling then requires $E_{J\alpha} \ll E_C=\min_n \{ E_n - E_{01}, E_n - E_{10} \}$.
Here, $E_{jk}$ stands for the charging energy of the state with $j$ ($k$) additional Cooper pairs on dot $a$ ($b$), and $n$ enumerates all charges states different from $(0,1)$ and $(1,0)$. 
 The corresponding effective Hamiltonian, acting in the reduced Hilbert space spanned by the charge states $(0,1)$ and $(1,0)$, can be constructed systematically order for order, cf.~\cite{cohen}. Carrying out this procedure up to order $\mathcal{O}(E_{J\alpha}/E_C)$ and discarding of phase-independent terms merely renormalizing the charging energies, we obtain
\begin{align}\label{heff1}
\hat{H}_\text{eff}=&\frac{h_0}{2}\hat\sigma_z- J( \hat{\sigma}^-e^{i\phi} + \hat{\sigma}^+e^{-i\phi})
 - J'(\hat{\sigma}^- + \hat{\sigma}^+).
\end{align}
Here, we have employed a pseudospin description for the charge states $(1,0)$ and $(0,1)$, reinterpreting them as the spin-up and spin-down eigenstates of the operator $\hat\sigma_z$. Any deviation from charge degeneracy (tunable by the gate voltages) leads to a ``magnetic field"  $h_0=E_{10}-E_{01}$. 

\begin{figure}
\centering
\includegraphics[width=1.0\columnwidth]{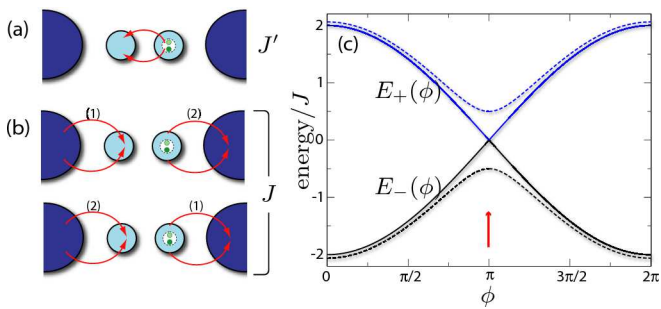} 
\caption{(color online) 
(a,b) Leading-order processes relevant to charge transfer in the three-junction system, close to a charge-degeneracy line. Depicted are the processes starting in the initial charge state $(0,1)$. The inter-dot tunneling (a) contributes to $J'$, the  dot-lead processes (b) contribute to $J$.
(c) Energies $E_\pm(\phi)$, Eq.\ \eqref{epm},  of the two-level system close to a charge-degeneracy line. The eigenenergies are plotted as functions of phase for the resonant case $J=J'$. The solid (dashed) lines correspond to $h_0=0$ ($J/2$), the arrow marks the position of the low-temperature I$\Phi$ discontinuity at $\pi$. 
\label{fig:fig3}}
\end{figure}

Transfer of Cooper pairs between dots and leads is described by the terms $\sim J,J'$ in Eq.\ \eqref{heff1}. Here, $J'$ originates from inter-dot tunneling processes, Fig.\ \ref{fig:fig3}(a), and $J$ is generated by dot-lead tunneling, Fig.\ \ref{fig:fig3}(b). In the spirit of the nomenclature for single electron tunneling processes, we may refer to these different processes as sequential tunneling and cotunneling of Cooper pairs. The energies $J,J'>0$  are given by
\begin{align}\label{Jeq}
J=\frac{E_{JL}E_{JR}}{8}&\sum_{j=0,1}\left[ \frac{1}{E_{jj}-E_{01}}+\frac{1}{E_{jj}-E_{10}}\right]
\end{align}
and $J'=E_{Jm}/2$.
 Note that the ratio $J/J'$ weakly depends on the gate voltages. Additional tunability of this parameter may be achieved by terminating the superconducting leads in a 2D electron gas, cf.\ e.g.\ \cite{1994PhRvB..50.8118M}, which enables the direct tuning of tunneling strengths.

\begin{figure}
\centering
\includegraphics[width=0.94\columnwidth]{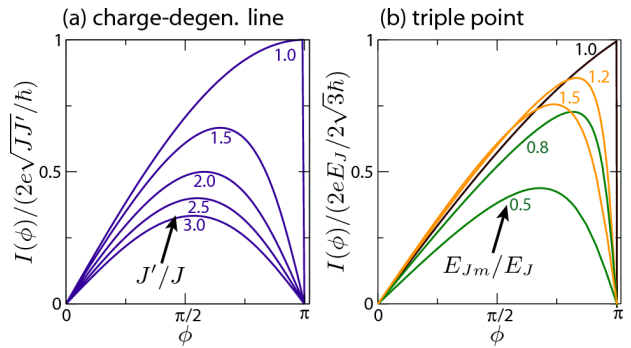} 
\caption{(color online) Current-phase relations for the three-junction JJA at zero temperature, (a) close to a charge-degeneracy line, (b) at a triple point. (a) I$\Phi$ for different ratios $J'/J$ of the inter-dot and dot-lead coupling. (b) I$\Phi$ for $E_J\equiv E_{JL}=E_{JR}$ and different ratios $E_{Jm}/E_J$.
\label{fig:fig4}}
\end{figure}

By diagonalization of Eq.\ \eqref{heff1} we obtain the eigenenergies plotted in Fig.\ \ref{fig:fig3}(c),
\be\label{epm}
E_\pm(\phi) =\pm\sqrt{(h_0/2)^2+\varepsilon_{JJ'}^2(\phi)},
\ee
where $\varepsilon_{JJ'}(\phi)=\left[J^2+(J')^2+2JJ'\cos\phi\right]^{1/2}$. Note that for $h_0=0$ the two states become degenerate at $\phi=\pi$. In the Heisenberg picture, the current is
\be\label{cr}
\hat{I} = \frac{2ie}{\hbar} J (\hat{\sigma}^-e^{i\phi} - \hat{\sigma}^+e^{-i\phi} ).
\ee
With this, the supercurrent can be computed for both zero and finite temperatures. For $T=0$, it is given by the ground state expectation value of the current operator,
\begin{align}\label{cphi}
I_0(\phi)
=\frac{2e}{\hbar}\frac{JJ'\sin\phi}{\sqrt{(h_0/2)^2+\varepsilon_{JJ'}^2(\phi)}}.
\end{align}
Alternatively, the zero-temperature supercurrent may be obtained via $I_0(\phi)=\frac{2e}{\hbar}\frac{\partial E_-}{\partial\phi}$, leading to the same result.
At finite temperatures small compared to $\Delta$, quasiparticle excitations may be neglected and the I$\Phi$ is obtained by averaging with respect to the equilibrium density matrix, resulting in $I(\phi)= I_0(\phi)\,
\tanh\left[ \beta E_-(\phi)\right]$.
This I$\Phi$ is depicted for $T=0$ in Fig.\ \ref{fig:fig4}(a). Note that due to the general symmetry properties of I$\Phi$s \cite{golubov} it is sufficient to plot the range $0\le\phi\le\pi$. This I$\Phi$ is intimately related to the result found in the single-grain case at charge-degeneracy points \cite{1993PhRvL..70.2940M}. In both cases, Cooper pair transfer involves the switching between two charge states -- occupation and de-occupation in the single-grain case, occupation of the right or the left grain in the present case. This analogy and the underlying pseudospin description are the origin for the I$\Phi$s to be \emph{structurally identical}. In particular, in close resemblance to the discontinuity for the single-grain system, we find that, given $J=J'$, the I$\Phi$ of the double-grain system undergoes a discontinuity at $\phi=\pi$,
\be\label{cphi2}
I_0(\phi)=\frac{2e}{\hbar}J\sin(\phi/2)\,\text{ for } -\pi<\phi<\pi,
\ee
 which gets broadened by temperature. However, similar to the situation of the sinusoidal relation in regime (i), the critical current at the charge-degeneracy line remains suppressed by a factor $E_{J\alpha}/E_C$, see Eq.\ \eqref{Jeq}. It is interesting to note that similar I$\Phi$ discontinuities are also present in classical and ballistic point contacts \cite{kulik,beenakker,golubov}.

A few comments on the I$\Phi$ \eqref{cphi} and its crossover to a sinusoid are in order. First, we remark that Eq.\ \eqref{cphi} is invariant under interchange $J\leftrightarrow J'$, implying a one-to-one correspondence between the supercurrent through a double-dot system with large inter-dot tunneling but small dot-lead coupling, and the current for small inter-dot tunneling but large dot-lead coupling. Second, the effect of detuning $h_0$ from the charge-degeneracy line can always be mimicked by adjusting the parameters $J$ and $J'$, so that the consideration of the $h_0=0$ case is sufficient. Third, for $J'/J\begin{smallmatrix}\gg\\\ll\end{smallmatrix}1$, we asymptotically recover a sinusoidal I$\Phi$,
\be
I(\phi)\simeq  \frac{2e}{\hbar}\min\{J,J'\}\sin\phi\, \tanh\big[\beta\max\{J,J'\}\big].
\ee
This is in accordance with the behavior of the single-grain I$\Phi$ when making the Josephson energies in the left and right junction very different \cite{1993PhRvL..70.2940M}. 

In the regime (iii), the system is tuned to a triple point of the charge-stability diagram, where three charge states become degenerate.   We focus on the triple point specified by a circle in Fig.\ \ref{fig:fig2}(b), where the corresponding Hamiltonian may be represented  as
\be
\hat{H}_\text{eff} = \frac{1}{2}\left(
\begin{array}{ccc}
2E_{00} & -E_{JL}e^{i\phi} & -E_{JR}e^{-i\phi}\\
-E_{JL}e^{-i\phi} & 2E_{10} & -E_{Jm}e^{-i\phi}\\
-E_{JR}e^{i\phi} & -E_{Jm}e^{i\phi} & 2E_{01}
\end{array}
\right).
\ee
For simplicity, we present an analytical I$\Phi$ expression only for $T=0$, exact degeneracy, and identical junction strengths.  Denoting $E_J\equiv E_{JL}=E_{JR}=E_{Jm}$, one finds
\be\label{trippt}
I_0(\phi)=\frac{2e}{\hbar}\frac{E_J}{3}\sin (\phi/3)\,\,\text{ for } -\pi < \phi < \pi,
\ee
giving the maximum possible supercurrent in the three-junction device.
Finite temperatures, detuning from the triple point, and differing junction strengths generally lead to a broadened I$\Phi$ with more complicated expressions. Eq.\ \eqref{trippt}, along with  examples for the situation of differing junction strengths, is plotted in Fig.\ \ref{fig:fig4}(b).

\emph{Generalizations.}---%
For an array consisting of $N$ junctions, maximally $N$ states can become degenerate at isolated points in the charge stability diagrams. The I$\Phi$ hierarchy then consists of $N$ different I$\Phi$s whose critical currents differ in powers of the small parameter $E_J/E_C$. Away from degeneracies, we find that the I$\Phi$ is sinusoidal with critical current suppressed by a factor $(E_J/E_C)^{N-1}$. In the vicinity of two-fold degeneracies, the I$\Phi$ takes the form Eq.\ \eqref{cphi2} with discontinuity at phase $\pi$; the critical current in this region is suppressed by $(E_J/E_C)^{N-2}$. When increasing the degree of degeneracy, this hierarchy continues and discontinuities persist.  

On the uppermost hierarchy level, the critical current is not suppressed and the discontinuity is fully developed when all junctions have the same Josephson energy $E_J$. In this case, the system can be mapped to an effective tight-binding model with periodic boundary conditions,
\be\label{htb}
\hat{H}_\text{tb}=-\frac{E_J}{2}\sum_\nu e^{-i\phi/N} \ket{\nu+1}\bra{\nu} +\text{H.c.}
\ee
Here, the states $\ket{\nu}$ denote the $N$ degenerate charge states and can be pictured as the states with one extra Cooper pair on one of the grains or the lead. The eigenenergies of the system are given by $\epsilon_n = -E_J\cos(k_n + \phi/N)$ with $k_n=2\pi n/N$ and $n=0,\cdots,(N-1)$.
The resulting zero-temperature I$\Phi$ is
\be\label{inew}
I_0^{(N)}(\phi)=\frac{2e}{\hbar}\frac{E_J}{N}\sin(\phi/N)\,\text{ for } -\pi<\phi<\pi,
\ee
which generalizes Eq.\ \eqref{trippt} for arbitrary number of junctions $N$, and approaches a shallow sawtooth for large $N$. Owing to the general $2\pi$ periodicity of I$\Phi$s \cite{golubov}, Eq.\ \eqref{inew} proves that discontinuities of the I$\Phi$ at $\pi$ persist.
It is interesting to note that Eq.\ \eqref{inew} is  reminiscent of the classical limit for a 1D JJA \cite{matveev2002}. The crucial difference is that, in the present case, there is a probability $1/N$ to occupy any of the charge states. As a result, the critical current scales with $N^{-2}$ instead of $N^{-1}$ for large $N$.

\emph{Conclusions.}---%
We have studied I$\Phi$s in finite-size 1D JJAs. In the limit $E_C\gg E_J$, the charge-stability diagram serves as an important tool to identify different types of I$\Phi$s. We have established a hierarchy of I$\Phi$s, where each hierarchy level belongs to a different degree of charge degeneracy and has  a distinct magnitude of the critical current. Under specific conditions and zero temperature, these I$\Phi$s display discontinuities at phase $\pi$, originating from a crossing of the lowest two eigenenergies. Maximum charge degeneracy allows for a mapping to a tight-binding model, and results in a peculiar I$\Phi$ converging to a sawtooth function with a critical current scaling with $N^{-2}$. With experiments on small \cite{geerligs,pashkin2003,bibow} and large arrays \cite{haviland} having been demonstrated, we believe that our predictions will have an experimental impact in the near future.

\begin{acknowledgments}
We thank E.\ Dupont, E.\ Thuneberg, L.\ Glazman, and M.\ Devoret for valuable discussions.  This work was supported  by NSF through the Yale Center for Quantum Information Physics and by Yale University.
\end{acknowledgments}


\end{document}